\documentclass[a4paper,12pt]{article}
\usepackage{amssymb,amsmath}
\usepackage{epsfig,fancyhdr}

\usepackage{a4}
\usepackage{parskip}
\usepackage{appendix}
\usepackage{color,soul,rotating}
\usepackage{amsthm,amsxtra,amscd,relsize,multirow}
\usepackage[all,2cell]{xy}\UseAllTwocells

\newcommand{\sect}[1]{\setcounter{equation}{0}\section{#1}}

\def\S{{\mathcal S}}

\def\J{{\mathcal J}}

\def\V{{\mathrm V}}

\def\sinh{\mathrm{sinh}}
\def\cosh{\mathrm{cosh}}

\def\axs{AdS_5\times S^5}
\newcommand{\eq}[1]{\begin{equation} #1 \end{equation}}
\newcommand{\al}[1]{\begin{align} #1 \end{align}}

\begin{document}
\begin{titlepage}
\markright{\bf TUW--11--08}
\title{Three- and four-point correlators of operators dual to folded string solutions in $\axs$}

\author{D.~Arnaudov${}^{\star}$, R.~C.~Rashkov${}^{\dagger,\star}$\thanks{e-mail:
rash@hep.itp.tuwien.ac.at.}\, and T.~Vetsov${}^{\star}$
\ \\ \ \\
${}^{\star}$  Department of Physics, Sofia
University,\\
5 J. Bourchier Blvd, 1164 Sofia, Bulgaria
\ \\ \ \\
${}^{\dagger}$ Institute for Theoretical Physics, \\ Vienna
University of Technology,\\
Wiedner Hauptstr. 8-10, 1040 Vienna, Austria
}
\date{}
\end{titlepage}

\maketitle
\thispagestyle{fancy}

\begin{abstract}
Recently there has been progress on the calculation of $n$-point correlation functions with two ``heavy'' (with large quantum numbers) states at strong coupling. We extend these findings by computing three-point functions corresponding to a folded three-spin semiclassical string with one angular momentum in AdS and two equal spins in the sphere. We recover previous results as limiting cases. Also, following a recent paper by Buchbinder and Tseytlin, we provide relevant four-point functions and consider some of their limits.
\end{abstract}

\sect{Introduction}

The attempts to establish a correspondence between the large $N$ limit of gauge theories and string theory has more than 30 years history and over the years it showed different faces. Recently an explicit realization of this correspondence was provided by the Maldacena conjecture about AdS/CFT correspondence \cite{Maldacena}. The convincing results from the duality between type IIB string theory on $AdS_5\times S^5$ and ${\cal N}=4$ super Yang-Mills theory~\cite{Maldacena,GKP,Witten} made this subject a major research area, and many fascinating new features have been established.

One of the consequences of this correspondence is that the planar correlators of single-trace conformal primary operators in the boundary gauge theory should be related to the correlation functions of the corresponding closed-string vertex operators on a worldsheet with $S^2$ topology. We begin by examining a correlator with two ``heavy'' vertex operators (with large quantum numbers of the order of the string tension) and a number of ``light'' vertex operators (with quantum numbers and dimensions of order one). Then the large $\sqrt{\lambda}$ behavior of correlation functions of such operators is fixed by a semiclassical string trajectory determined by the ``heavy'' operator insertions, and with sources provided by the vertex operators of the ``light'' states. The exact procedure is the following. First, we construct the classical string solution that determines the leading large $\sqrt{\lambda}$ contribution to the correlator of the ``heavy'' operators. Second, we calculate the full correlation function by evaluating the product of ``light'' vertex operators on this solution.

This semiclassical approach was developed for the computation of two-point functions in \cite{Polyakov:2002}-\cite{Buchbinder:2010vw}. A generalization to certain three-point functions was discussed in \cite{Janik:2010gc,Buchbinder:2010vw}, and addressed in \cite{Zarembo:2010,Costa:2010}, where the ``heavy'' operators corresponded to a semiclassical string state with large spin in $S^5$ and the ``light'' operator represented a BPS state -- massless (supergravity) scalar or dilaton mode. The extension to vertex operator insertions was made in \cite{Roiban:2010}. Further developments can be traced in subsequent papers~\cite{Hernandez:2010}-\cite{Park:2010}. Recently, a generalization to four-point functions was initiated in \cite{BT:2010}, and various correlators were calculated. Furthermore, the validity of the method was checked via differentiation by $\lambda$, and comparison with results from the gauge theory side was provided.

Motivated by these studies, we consider the 3-point correlation functions of two ``heavy'' operators, corresponding to a folded string solution with three spins (one in $AdS_5$ and two equal ones in $S^5$), and one BPS (dilaton or chiral primary) operator. We generalize some of the results presented in \cite{Roiban:2010}. Furthermore, we extend the findings of~\cite{BT:2010} by calculating four-point functions of two ``heavy'' operators, which correspond to solutions with two and three spins, and two BPS operators.

The paper is organized as follows. To explain the method, in the next Section we give a short review of the procedure for calculating semiclassically $n$-point correlation functions via vertex operators. Next, we proceed with the calculation of three-point correlation functions for a particular three-spin string solution, taking either the dilaton or the chiral primary operator (CPO) as the ``light'' operator. We provide several limiting cases of the correlators. Then, we compute a number of related four-point functions and their limits. We conclude with a brief discussion on the results.

\sect{Correlation functions with two ``heavy'' operators}

Let us start with reviewing the case of two-point correlators. Their calculation in the leading semiclassical approximation is intimately related to finding an appropriate classical string solution \cite{Tseytlin:2003}-\cite{Janik:2010gc}. If $V_{H1}(\xi_1)$ and $V_{H2}(\xi_2)$ are the two ``heavy'' vertex operators inserted at points $\xi_1$ and $\xi_2$ on the worldsheet (chosen as a plane or a sphere, because we consider planar AdS/CFT duality), the two-point function for large string tension ($\sqrt{\lambda}\gg1$) is determined by the stationary point of the action
\eq{
\langle V_{H1}(\xi_1)V_{H2}(\xi_2)\rangle\sim e^{-I},
}
where $I$ is the action of the $\axs$ superstring sigma model in the embedding coordinates
\al{
&I=\frac{\sqrt{\lambda}}{4\pi}\int d^2\xi\ \Big(\partial Y_M\bar{\partial}Y^M + \partial X_k\bar{\partial}X_k + {\rm fermions}\Big)\,,\\
&Y_MY^M=-Y_0^2-Y_5^2+Y_1^2+Y_2^2+Y_3^2+Y_4^2=-1\,,\quad\ X_kX_k=X_1^2+\ldots+X_6^2=1\,.\nonumber
}
We work in conformal gauge and use a worldsheet with Euclidean signature. Thus, the 2D derivatives are $\partial=\partial_1+i\partial_2,\,\bar{\partial}=\partial_1-i\partial_2$. The relation between the embedding coordinates, and the global and Poincar\'e coordinates in $AdS_5$ that we will need below is
\al{
&Y_5+iY_0=\cosh\,\rho\ e^{it},\quad
Y_1+iY_2=\sinh\,\rho\,\cos\theta\,e^{i\phi_1},\quad
Y_3+iY_4=\sinh\,\rho\,\sin\theta\,e^{i\phi_2},\nonumber\\
&Y_m=\frac{x_m}{z}\,,\qquad
Y_4=\frac{1}{2z}(-1+z^2+x^mx_m)\,,\qquad
Y_5=\frac{1}{2z}(1+z^2+x^m x_m)\,,
\label{poincare}
}
where $x^mx_m=-x_0^2+x_ix_i\ (m=0,1,2,3;\ i=1,2,3)$. However, for convenience we will use from now on the Euclidean continuation of $AdS_5$
\eq{
t_e=it\,,\qquad Y_{0e}=iY_0\,,\qquad x_{0e}=ix_0\,,
}
so that $Y_MY^M=-Y_5^2+Y_{0e}^2+Y_iY_i+Y_4^2=-1$.

The stationary point solution solves the string equations of motion with singular sources determined by $V_{H1}(\xi_1)$ and $V_{H2}(\xi_2)$. Using the conformal symmetry of the theory one can map the $\xi$-plane to the Euclidean cylinder parameterized by $(\tau_e,\sigma)$
\eq{
e^{\tau_e+i\sigma}=\frac{\xi-\xi_2}{\xi-\xi_1}\,.
\label{confmap}
}
Under this conformal map the singular solution on the $\xi$-plane transforms into a smooth classical string solution on the cylinder \cite{Tseytlin:2003,Buchbinder:2010,Buchbinder:2010vw}. The new solution carries the same quantum numbers as the states represented by the vertex operators, so that no information is lost.

The discussion above can be repeated for a physical integrated vertex operator labeled by a point ${\rm x}$ on the boundary of the Poincar\'e patch of $AdS_5$ \cite{Polyakov:2002,Tseytlin:2003}
\eq{
{\rm V}_H({\rm x})=\int d^2\xi\ V_H(\xi;{\rm x})\,,\qquad V_H(\xi;{\rm x})\equiv V_H(z(\xi),x(\xi)-{\rm x},X_k(\xi))\,.
}
The semiclassical two-point correlator $\langle\V_{H1}({\rm x}_1)\V_{H2}({\rm x}_2)\rangle$ is again fixed by the classical action on the stationary point solution. After we apply the conformal map to the cylinder~\eqref{confmap} we obtain a smooth solution that is actually the corresponding spinning string solution in terms of Poincar\'e coordinates which satisfies the boundary conditions (see \cite{Buchbinder:2010vw} for details)
\eq{
\tau_e\rightarrow-\infty\ \ \Longrightarrow\ \ z\rightarrow0\,,\ \ x\rightarrow{\rm x}_1\,,\qquad\qquad\tau_e\rightarrow+\infty\ \ \Longrightarrow\ \ z\rightarrow0\,,\ \ x\rightarrow{\rm x}_2\,.
\label{boundcond}
}

In a similar way one can calculate semiclassically three-point functions with two ``heavy'' and one ``light'' operators \cite{Zarembo:2010,Roiban:2010}
\al{
G_3({\rm x}_1,{\rm x}_2,{\rm x}_3)&=\langle\V_{H1}({\rm x}_1)\V_{H2}({\rm x}_2)\V_L({\rm x}_3)\rangle\\
&=\int{\cal D}\mathbb{X}^\mathbb{M}\ e^{-I}\int d^2\xi_1d^2\xi_2d^2\xi_3\ V_{H1}(\xi_1;{\rm x}_1)V_{H2}(\xi_2;{\rm x}_2)V_L(\xi_3;{\rm x}_3)\,,\nonumber
}
where $\int{\cal D}\mathbb{X}^\mathbb{M}$ is the integral over $(Y_M,X_k)$. In the stationary point equations the contribution of the ``light'' operator can be neglected, so that the solution is the same as in the case of two-point function of two ``heavy'' operators. Thus we obtain \cite{Roiban:2010}
\eq{
\frac{G_3({\rm x}_1,{\rm x}_2,{\rm x}_3)}{G_2({\rm x}_1,{\rm x}_2)}=\int d^2\xi\ V_L(z(\xi),x(\xi)-{\rm x}_3,X_k(\xi))\,,
\label{strconstxi}
}
where $(z(\xi),x(\xi),X_k(\xi))$ represents the corresponding string solution with the same quantum numbers as the ``heavy'' vertex operators, and with the boundary conditions \eqref{boundcond} transformed to the $\xi$-plane by \eqref{confmap}. With the help of the 2D conformal invariance we can also represent \eqref{strconstxi} in terms of the cylinder ($\int d^2\sigma=\int^\infty_{-\infty}d\tau_e\int^{2\pi}_0d\sigma$)
\eq{
\frac{G_3({\rm x}_1,{\rm x}_2,{\rm x}_3)}{G_2({\rm x}_1,{\rm x}_2)}=\int d^2\sigma\ V_L(z(\tau_e,\sigma),x(\tau_e,\sigma)-{\rm x}_3,X_k(\tau_e,\sigma))\,.
\label{strconstst}
}

The global conformal $SO(2,4)$ symmetry determines the two- and three-point correlators (assuming that $\V_{H2}=\V^*_{H1}$)
\al{\label{2point}
G_2({\rm x}_1,{\rm x}_2)&=\frac{C_{12}\ \delta_{\Delta_1\!,\Delta_2}}{{\rm x}_{12}^{\Delta_1+\Delta_2}}\,,\qquad{\rm x}_{ij}\equiv|{\rm x}_i-{\rm x}_j|\,,\\
G_3({\rm x}_1,{\rm x}_2,{\rm x}_3)&=\frac{C_{123}}{{\rm x}_{12}^{\Delta_1+\Delta_2-\Delta_3}{\rm x}_{13}^{\Delta_1+\Delta_3-\Delta_2}
{\rm x}_{23}^{\Delta_2+\Delta_3-\Delta_1}}\,,
\label{3point}
}
where $\Delta_i$ are the dimensions of the operators. With a proper choice of ${\rm x}_i$ one can remove the dependence on ${\rm x}_{ij}$ in \eqref{strconstst}, and use \eqref{strconstst} to calculate the structure constants $C_{123}$~\cite{Zarembo:2010,Roiban:2010}. Presuming that $\Delta_1=\Delta_2$, we find (after setting $C_{12}=1$ in \eqref{2point}) that
\eq{
\frac{G_3({\rm x}_1,{\rm x}_2,{\rm x}_3=0)}{G_2({\rm x}_1,{\rm x}_2)}=C_{123}\left(\frac{{\rm x}_{12}}{|{\rm x}_1|\,|{\rm x}_2|}\right)^{\Delta_3}\!.
\label{norm3point}
}

Now let us concentrate on the case of the four-point function
\al{
&G_4({\rm x}_1,{\rm x}_2,{\rm x}_3,{\rm x}_4)=\langle\V_{H1}({\rm x}_1)\V_{H2}({\rm x}_2)\V_{L1}({\rm x}_3)\V_{L2}({\rm x}_4)\rangle\\
&=\int{\cal D}\mathbb{X}^\mathbb{M}\ e^{-I}\int d^2\xi_1d^2\xi_2d^2\xi_3d^2\xi_4\ V_{H1}(\xi_1;{\rm x}_1)V_{H2}(\xi_2;{\rm x}_2)
V_{L1}(\xi_3;{\rm x}_3)V_{L2}(\xi_4;{\rm x}_4)\,.\nonumber
}
The semiclassical trajectory being the same, to compute the leading term in $G_4$ we need to evaluate the product of the ``light'' operators on the solution
\eq{
\frac{G_4({\rm x}_1,{\rm x}_2,{\rm x}_3,{\rm x}_4)}{G_2({\rm x}_1,{\rm x}_2)}=\!\int\!d^2\xi_3\,V_{L1}(z(\xi_3),x(\xi_3)-{\rm x}_3,X_k(\xi_3))\!\!\int\!d^2\xi_4\,V_{L2}(z(\xi_4),x(\xi_4)-{\rm x}_4,X_k(\xi_4))\,.
}
Transforming to the $(\tau_e,\sigma)$-coordinates we obtain
\al{
\frac{G_4({\rm x}_1,{\rm x}_2,{\rm x}_3,{\rm x}_4)}{G_2({\rm x}_1,{\rm x}_2)}&=
\int d^2\sigma\ V_{L1}(z(\tau_e,\sigma),x(\tau_e,\sigma)-{\rm x}_3,X_k(\tau_e,\sigma))\nonumber\\
&\times\int d^2\sigma'\ V_{L2}(z(\tau'_e,\sigma'),x(\tau'_e,\sigma')-{\rm x}_4,X_k(\tau'_e,\sigma'))\,.
\label{4point}
}
From \eqref{strconstst} and \eqref{4point} we get the factorization
\al{
\langle\V_{H1}({\rm x}_1)\V_{H2}({\rm x}_2)\V_{L1}({\rm x}_3)\V_{L2}({\rm x}_4)\rangle&=
\frac{\langle\V_{H1}({\rm x}_1)\V_{H2}({\rm x}_2)\V_{L1}({\rm x}_3)\rangle\langle\V_{H1}({\rm x}_1)\V_{H2}({\rm x}_2)\V_{L2}({\rm x}_4)\rangle}
{\langle\V_{H1}({\rm x}_1)\V_{H2}({\rm x}_2)\rangle}\,,\nonumber\\
G_4({\rm x}_1,{\rm x}_2,{\rm x}_3,{\rm x}_4)&=\frac{G_3({\rm x}_1,{\rm x}_2,{\rm x}_3)\,G_3({\rm x}_1,{\rm x}_2,{\rm x}_4)}{G_2({\rm x}_1,{\rm x}_2)}\,.
\label{factorization}
}

For more details we refer to \cite{Buchbinder:2010,Buchbinder:2010vw,Roiban:2010,BT:2010}.

\sect{Three-point correlators for a folded string solution with three spins in $\axs$}

In this Section we apply the methods described above to the calculation of particular three-point functions. Without loss of generality we choose ${\rm x}_1=(-1,0,0,0)$ and ${\rm x}_2=(1,0,0,0)$. We consider a generalization of (3.14) in \cite{Roiban:2010} for the string solution that determines the semiclassical trajectory. Our solution has a large spin $S=\sqrt{\lambda}\,{\cal S}$ in AdS and two orbital momenta $J_1=J_2$ in $S^5$, and is defined as
\al{
\label{sol1}
t_e&=\kappa\tau_e\,,\quad\phi_1=-i\kappa\tau_e\,,\quad\rho=\mu\sigma\,,\quad\mu\approx\frac{1}{\pi}\ln{\cal S}\gg1\,,\quad\kappa=\sqrt{\mu^2+\nu^2+n^2}\,,\\
\label{sol2}
\gamma&=\frac{\pi}{2}\,,\quad\psi=n\sigma\,,\quad\varphi_1=\varphi_2=-i\nu\tau_e\,,\quad\nu=\J=\frac{J}{\sqrt{\lambda}}\,,\quad J=J_1+J_2\,,
}
where $(\gamma,\psi,\varphi_1,\varphi_2,\varphi_3)$ are the coordinates of $S^5$. The background \eqref{sol1} approximates the exact elliptic function solution\footnote{See \cite{GKP2} for the particular case of $\nu,n=0$.} in the limit $\kappa,\mu\gg1$ on the interval $\sigma\in[0,\frac{\pi}{2}]$. To obtain the formal periodic solution on $0<\sigma\leq2\pi$ one needs to combine together four stretches $\rho=\mu\sigma$ of the closed, and folded onto itself, string.

In the embedding coordinates the AdS part of the solution is
\al{
Y_5&=\cosh(\kappa\tau_e)\,\cosh(\mu\sigma)\,,\qquad Y_{0e}=\sinh(\kappa\tau_e)\,\cosh(\mu\sigma)\,,\qquad Y_4=0\,,\nonumber\\
Y_1&=\cosh(\kappa\tau_e)\,\sinh(\mu\sigma)\,,\qquad Y_2=-i\,\sinh(\kappa\tau_e)\,\sinh(\mu\sigma)\,,\qquad Y_3=0\,.
}
Alternatively, in Poincar\'e coordinates \eqref{poincare}
\al{
z&=\frac{1}{\cosh(\kappa\tau_e)\,\cosh(\mu\sigma)}\,,\qquad x_{0e}=\tanh(\kappa\tau_e)\,,\qquad x_1=\tanh(\mu\sigma)\,,\\
x_2&=-i\tanh(\kappa\tau_e)\tanh(\mu\sigma)\,,\qquad x_3=0\,,\qquad z^2+x_{0e}^2+x_1^2+x_2^2+x_3^2=1\,.
}

The energy of the solution is
\eq{
E-S=\sqrt{J^2+\frac{\lambda}{\pi^2}\ln^2{\cal S}+\lambda n^2}=\frac{\sqrt{\lambda}}{\pi}\sqrt{1+\ell_1^2+\ell_2^2}\,\ln{\cal S}\,,\quad\ell_1\equiv\frac{\nu}{\mu}\,,\quad\ell_2\equiv\frac{n}{\mu}\,.
}

Now we will examine the three-point functions with two ``heavy'' operators and one ``light'' operator, which will be either the dilaton, or the superconformal primary scalar (chiral primary operator).

\subsection{Dilaton as ``light'' operator}

The 10D dilaton field is decoupled from the metric in the Einstein frame \cite{Kim:1985}. Thus, it satisfies the free massless 10D Laplace equation in $\axs$. The corresponding string vertex operator is proportional to the worldsheet Lagrangian (we will keep nonzero value~$j$ of $S^5$ momentum)
\al{\label{dilvertex}
V_L({\rm x}=0)&=V^{(\rm dil)}_j(0)=\hat{c}_{\Delta}K_{\Delta}\,{\rm X}^j\,\big(\partial Y_M\bar{\partial}Y^M + \partial X_k\bar{\partial}X_k + {\rm fermions}\big)\,,\\
K_{\Delta}&\equiv\left(\frac{z}{z^2+x^mx_m}\right)^{\Delta}\!,\qquad{\rm X}\equiv X_1+iX_2=\sin\gamma\cos\psi\,e^{i\varphi_1},\nonumber
}
where $\hat{c}_\Delta$ depends only on the normalization of the dilaton. Here and below we will ignore the fermionic terms and overall normalization factors in the vertex operators. To the leading order in the large $\sqrt{\lambda}$ expansion $\Delta=4+j$. The respective dual gauge theory operator is proportional to ${\rm tr}(F^2_{mn}Z^j+\ldots)$ or, for $j=0$, just the SYM Lagrangian.

From \eqref{strconstst}, \eqref{norm3point} and \eqref{dilvertex} follows that
\al{
\label{dilstrconst}
C_{123}&=c_{\Delta}\int^{\infty}_{-\infty}d\tau_e\int^{2\pi}_0d\sigma\,K_{\Delta}\,U\,,\qquad c_{\Delta}=2^{-\Delta}\hat{c}_{\Delta}\,,\\
U&={\rm X}^j\Big[z^{-2}(\partial x_m\bar{\partial}x^m+\partial z\bar{\partial}z) + \partial X_k\bar{\partial}X_k\Big].
}
The normalization constant $\hat{c}_\Delta$ of the dilaton vertex operator was calculated in \cite{Roiban:2010}
\eq{
\hat{c}_\Delta=\hat{c}_{4+j}=\frac{\sqrt{\lambda}}{8\pi N}\sqrt{(j+1)(j+2)(j+3)}\,.
}

Evaluating $U$ on the large-spin folded string classical solution in \eqref{sol1} and \eqref{sol2} we obtain
\eq{
U=e^{j\nu\tau_e}\cos^j(n\sigma)\,(\kappa^2\cosh^2\rho+\mu^2-\kappa^2\sinh^2\rho-\nu^2+n^2)=2(\mu^2+n^2)\,e^{j\nu\tau_e}\cos^j(n\sigma)\,,
}
so that the expression in \eqref{dilstrconst} becomes
\eq{
C_{123}=4c_\Delta\int^{\infty}_{-\infty}d\tau_e\int^{\pi/2}_0d\sigma\,\frac{2(\mu^2+n^2)\,e^{j\nu\tau_e}\cos^j(n\sigma)}
{\left[\cosh(\kappa\tau_e)\,\cosh(\mu\sigma)\right]^{4+j}}\,,
}
where we have used that the solution for $\rho$ in \eqref{sol1} approximates the exact folded solution for $\mu\gg1$. Computing the structure constant, we get
\al{
C_{123}&=8c_{\Delta}2^{-j}\,\frac{\mu^2+n^2}{\mu\kappa}\,{\cal I}_{\tau}\sum_{k=0}^j\left(\!\begin{array}{c} j \\ k \end{array}\!\right){\cal I}_{\sigma}\,,\\
{\cal I}_{\tau}&=2^{4+j}\!\left[\frac{_2F_1\big(4+j,\frac{b_+}{2};1+\frac{b_+}{2};-1\big)}{b_+}+ \frac{_2F_1\big(4+j,\frac{b_-}{2};1+\frac{b_-}{2};-1\big)}{b_-}\right]\!,\\
{\cal I}_{\sigma}&=\frac{2^{4+j}}{\delta}\Big[e^{\pi\mu\delta/2}\ _2F_1\big(4+j,{\textstyle\frac{\delta}{2}};1+{\textstyle\frac{\delta}{2}};-e^{\pi\mu}\big)- {_2F_1}\big(4+j,{\textstyle\frac{\delta}{2}};1+{\textstyle\frac{\delta}{2}};-1\big)\Big],\\
b_{\pm}&\equiv4+j\big(1\pm{\textstyle\frac{\nu}{\kappa}}\big),\qquad\delta\equiv4+j+i(2k-j){\textstyle\frac{n}{\mu}}\,.\nonumber
}

Taking various limits of the above expression is fairly non-trivial, especially with respect to $\mu$. Around $n=0$ we have
\eq{
\textstyle{\cal I}_{\sigma}=\sinh\big(\frac{\pi\mu}{2}\big)\ _2F_1\big(\frac12,\frac{5+j}{2};\frac32;-\sinh^2\big(\frac{\pi\mu}{2}\big)\big)+{\rm O}(n^2)\,,
}
which means that the structure constant indeed reduces to (4.10) in \cite{Roiban:2010}. The limit $\mu\rightarrow0$ yields
\al{\nonumber
C_{123}&\approx8c_{\Delta}2^{-j}\,\frac{n}{\sqrt{\nu^2+n^2}}\,{{\cal I}_{\tau}}_{|\kappa=\sqrt{\nu^2+n^2}}
\sum_{k=0}^j\left(\!\begin{array}{c} j \\ k \end{array}\!\right)\frac{\sin\!\big(\frac{(2k-j)n\pi}{2}\big)}{2k-j}\\ \label{dilmu0}
&=8c_{\Delta}2^{-j}\,\frac{n}{\sqrt{\nu^2+n^2}}\,{{\cal I}_{\tau}}_{|\kappa=\sqrt{\nu^2+n^2}}\,\frac{ie^{-ijn\pi/2}}{2j}\\
&\times\!\Big[\textstyle{_2F_1}\big(-j,-\frac{j}{2};1-\frac{j}{2};-e^{in\pi}\big)
-e^{ijn\pi}\ {_2F_1}\big(-j,-\frac{j}{2};1-\frac{j}{2};-e^{-in\pi}\big)\Big].
\nonumber
}
It can be seen that the 3-point function is nonvanishing in this limit, because we need also $n\rightarrow0$ to have a geodesic for the classical trajectory.

For the case $j=0$ we obtain
\eq{
C_{123}=\frac{64c_{\Delta}(\S-1)(\S^2+4\S+1)\Big(\ln\S+\frac{\pi^2n^2}{\ln\S}\Big)}{9\pi(\S+1)^3\,\sqrt{\J^2+\frac{\ln^2\S}{\pi^2}+n^2}}\,.
}
In the large ${\cal S}$ limit, conforming to the discussion in \cite{Roiban:2010}, we get
\eq{
C_{123}\sim\frac{\ln S}{\sqrt{J^2+\frac{\lambda\ln^2S}{\pi^2}+\lambda n^2}}\,.
}

Yet another limit is taking $j\rightarrow\infty$ and ${\cal S}\rightarrow\infty$, while keeping $\ell_1$ and $\ell_2$ constant. Then the integrals over $\tau_e$ and $\sigma$ can be both evaluated with a saddle-point approximation
\al{\label{diljinf}
C_{123}&\approx\frac{8\pi c_{\Delta}e^{jh(\ell_1,\,\ell_2)}}{j}\,,\\
h(\ell_1,\ell_2)&=-\frac12\left[\ln\frac{1+\ell_1^2+\ell_2^2}{1+\ell_2^2}+\frac{\ell_1}{\sqrt{1+\ell_1^2+\ell_2^2}}\ln\frac{\sqrt{1+\ell_1^2+\ell_2^2}-\ell_1}
{\sqrt{1+\ell_1^2+\ell_2^2}+\ell_1}\right],
\label{diljinf1}
}
where
\eq{
h(\ell_1,\ell_2)=\left\{
\begin{array}{ll}
\frac{\ell_1^2}{2(1+\ell_2^2)}-\frac{5\ell_1^4}{12(1+\ell_2^2)^2}+\ldots&,\ \ \ \ \ell_1\rightarrow0
\\
\ln2-\frac12(1+\ell_2^2)\Big(\frac12+\ln\frac{2}{\sqrt{1+\ell_2^2}}+\ln\ell_1\Big)\frac{1}{\ell_1^2}+\ldots&,\ \ \ \ \ell_1\rightarrow\infty
\end{array}
\right.
\ ,
\label{diljinf2}
}
which should go to the results obtained in \cite{Roiban:2010} for $\ell_2=0$.

\subsection{Superconformal primary scalar as ``light'' operator}

The massless string state corresponding to the chiral primary operator originates from the trace of the graviton in $S^5$ directions which induces also the components of the graviton in $AdS_5$ directions, and mixes with the RR 5-form \cite{Kim:1985,Lee}. As discussed in \cite{Zarembo:2010,Berenstein:1998}, the bosonic part of the respective vertex operator assumes the following form (we ignore derivative terms that will not influence the calculation done below since we have chosen ${\rm x}_1=-{\rm x}_2$)\footnote{See \cite{BT:2010} for details.}
\al{\label{cpovertex}
V_L({\rm x}=0)&=V^{(\rm CPO)}_j(0)=\hat{c}_{\Delta}K_{\Delta}\,{\rm X}^j\big[z^{-2}(\partial x_m\bar{\partial}x^m-\partial z\bar{\partial}z) - \partial X_k\bar{\partial}X_k\big]\,,\\
K_{\Delta}&\equiv\left(\frac{z}{z^2+x^mx_m}\right)^{\Delta}\!,\qquad{\rm X}\equiv X_1+iX_2=\sin\gamma\cos\psi\,e^{i\varphi_1},\nonumber
}
where again $\hat{c}_\Delta$ depends only on the normalization of the scalar. The dual gauge theory operator is the BMN operator ${\rm tr}Z^j$ with dimension $\Delta=j$.

It can be inferred from \eqref{strconstst}, \eqref{norm3point} and \eqref{cpovertex} that
\al{
\label{cpostrconst}
C_{123}&=c_{\Delta}\int^{\infty}_{-\infty}d\tau_e\int^{2\pi}_0d\sigma\,K_{\Delta}\,U\,,\qquad c_{\Delta}=2^{-\Delta}\hat{c}_{\Delta}\,,\\
U&={\rm X}^j\Big[z^{-2}(\partial x_m\bar{\partial}x^m-\partial z\bar{\partial}z) - \partial X_k\bar{\partial}X_k\Big]\,,
}
where the normalization constant $\hat{c}_\Delta$ of the CPO is \cite{Zarembo:2010,Berenstein:1998}
\eq{
\hat{c}_\Delta=\hat{c}_j=\frac{\sqrt{\lambda}}{8\pi N}(j+1)\sqrt{j}\,.
}

We evaluate $U$ on the solution in \eqref{sol1} and \eqref{sol2}, and obtain
\eq{
U=2e^{j\nu\tau_e}\cos^j(n\sigma)\!\left(\frac{\kappa^2}{\cosh^2(\kappa\tau_e)}-\mu^2\tanh^2(\mu\sigma)-n^2\!\right)\!,
}
so that \eqref{cpostrconst} assumes the following form
\eq{
C_{123}=8c_j\int^{\infty}_{-\infty}d\tau_e\int^{\pi/2}_0d\sigma\,\frac{e^{j\nu\tau_e}\cos^j(n\sigma)}
{\big[\cosh(\kappa\tau_e)\,\cosh(\mu\sigma)\big]^j}\left(\frac{\kappa^2}{\cosh^2(\kappa\tau_e)}-\mu^2\tanh^2(\mu\sigma)-n^2\!\right)\!.
\label{cpostrconst'}
}
The result for the integral is too complicated to be presented here, so we will only provide various limits of the structure constant. We note that in contrast to the dilaton case taking the limit $n\rightarrow0$ would not render the findings of \cite{Roiban:2010} because this limit does not commute in general with the integrations in \eqref{cpostrconst'}. Consequently, by taking this limit we obtain original results.

First, let us take $\ell_1\rightarrow\infty$ and $\ell_2\rightarrow0$. We get for the structure constant
\al{
C_{123}&=c_j\,\frac{2^j}{j^2}\left(
8+\frac{\sqrt{\pi}(2-5j)j\Gamma[\frac{j}{2}]}{\Gamma[\frac{3+j}{2}]}+\frac{8j}{4+j}\ {_2F_1}\big(\textstyle1,1-\frac{j}{2};3+\frac{j}{2};-1\big)\right.\\
&+\left.\frac{16j\tanh\!\big(\frac{\pi\mu}{2}\big)}{(1+j)\,\cosh^j\big(\frac{\pi\mu}{2}\big)}-
2^{4+j}\frac{1-j}{1+j}\,e^{\pi\mu j/2}\ {_2F_1}\big(\textstyle j,\frac{j}{2};1+\frac{j}{2};-e^{\pi\mu}\big)\right)\ell_1+{\rm O}(\ell_1^{-1})\,.
\nonumber
}
Second, we will consider the limit $\ell_1\rightarrow0$ and $\ell_2\rightarrow0$, while taking ${\cal S}=e^{\pi\mu}\rightarrow\infty$
\eq{
C_{123}\approx\frac{\pi c_j\Gamma[\frac{j}{2}]}{\Gamma[\frac{1+j}{2}]}\!
\left[\frac{4}{\sqrt{\pi}}\!\left(\frac{1}{j}+\frac{{_2F_1}(1,1-\frac{j}{2};3+\frac{j}{2};-1)}{4+j}\right)
-\frac{3(2-j)\Gamma[\frac{j}{2}]}{(1+j)\Gamma[\frac{1+j}{2}]}\right].
\label{cpomuinfl10}
}
Third, let us take the small $\mu$ limit. We obtain a result that is still complicated, so we take $\ell_2\rightarrow0$ and get for $\ell_1\rightarrow0$
\eq{
C_{123}\approx\frac{8\pi^{3/2}c_j\Gamma[1+\frac{j}{2}]\mu}{(1+j)\Gamma[\frac{1+j}{2}]}\,,
}
which coincides with (4.30) in \cite{Roiban:2010}. Alternatively, for $\ell_1\rightarrow\infty$ we obtain
\eq{
C_{123}\approx\frac{8\pi c_j2^j}{1+j}\,\mu\ell_1\,,
\label{cpomu0l1inf}
}
which again is the same as (4.31) in \cite{Roiban:2010}. Thus, we can also recover the result \cite{Lee} for the three-point correlator of three BMN-type operators.

\sect{Four-point functions}

In this Section we will consider four-point correlators with two ``heavy'' operators corresponding to the folded string solutions (3.14) of \cite{Roiban:2010}, and \eqref{sol1}-\eqref{sol2} in the present paper, albeit for generic positions of the ``heavy'' vertex operators. We will follow closely the calculations done in \cite{BT:2010}.

\subsection{Folded string with one spin in $AdS_5$ and one in $S^5$}

The large spin gauge theory operators in the functions we will consider are dual to a folded string with spin $S$ in $AdS_5$ and $J$ in $S^5$ \cite{Roiban:2010}. Nevertheless, compared to \cite{Roiban:2010} here we start with generic positions ${\rm x}_1,\,{\rm x}_2$ of the form ${\rm x}_1=-{\rm x}_2$. The corresponding Euclidean stationary point solution is (we can choose the points ${\rm x}_1$ and ${\rm x}_2$ to lie on the $x_{0e}$-axis)
\al{\nonumber
&z=\frac{{\rm x}_{12}}{2\cosh(\kappa\tau_e)\,\cosh(\mu\sigma)}\,,\qquad x_{0e}=\frac{{\rm x}_{12}}{2}\tanh(\kappa\tau_e)+\frac12({\rm x}_1+{\rm x}_2)\,,\\ \label{x12sol}
&x_1+ix_2=\frac{{\rm x}_{12}}{2}\frac{\tanh(\mu\sigma)}{\cosh(\kappa\tau_e)}\,e^{i\phi_1},\qquad\phi_1=-i\kappa\tau_e\,,\qquad\varphi_1=-i\nu\tau_e\,,\\
&\kappa=\sqrt{\mu^2+\nu^2}\,,\qquad\mu=\frac{1}{\pi}\ln{\cal S}\,,\qquad{\cal S}=\frac{S}{\sqrt{\lambda}}\gg1\,,\qquad\nu={\cal J}=\frac{J}{\sqrt{\lambda}}\,.
\nonumber
}
Ignoring for now any contribution from ``light'' operators, we denote the respective ``heavy'' vertex operator as $\V_{S,-J}$, with $\V_{-S,J}\equiv\V^*_{S,-J}$. The two-point function of such operators can be calculated in the limit of large ${\cal S}$ \cite{Buchbinder:2010,Buchbinder:2010vw,Roiban:2010}
\eq{
\langle\V_{S,-J}({\rm x}_1)\V_{-S,J}({\rm x}_2)\rangle=\frac{1}{{\rm x}_{12}^{2\Delta(S,J)}}\,,\qquad
\Delta(S,J)=S+\sqrt{J^2+\frac{\sqrt{\lambda}}{\pi}\ln\frac{S}{\sqrt{\lambda}}}+\ldots\,.
\label{2pointx12}
}

The corresponding three-point functions with two large spin and one BPS operators have already been computed in \cite{Roiban:2010} for specific values of ${\rm x}_1$ and ${\rm x}_2$. We need 3-point correlators for general ${\rm x}_1$ and ${\rm x}_2$, and thus we will provide them below. We will consider three choices for the two ``light'' operators in the four-point functions, namely dilaton-dilaton, dilaton-CPO (the superconformal primary scalar), and CPO-CPO.

\paragraph{Dilaton-dilaton}\

The three-point function with one dilaton assumes the following form
\eq{
G_3({\rm x}_1,{\rm x}_2,{\rm x}_3=0)=\langle\V_{S,-J}({\rm x}_1)\V_{-S,J-j}({\rm x}_2)\V_j(0)\rangle\,,\qquad S,J\gg j\,,
\label{3dilaton}
}
where $j$ is the angular momentum of the dilaton, and we have explicitly maintained the $S^5$ angular momentum conservation. The structure constants are then determined by the ratio $G_3({\rm x}_1,{\rm x}_2,{\rm x}_3=0)/G_2({\rm x}_1,{\rm x}_2)$, which is given by the integral of the vertex operator~\eqref{dilvertex}. Evaluated on \eqref{x12sol}, it can be expressed as
\eq{
\frac{G_3({\rm x}_1,{\rm x}_2,{\rm x}_3=0)}{G_2({\rm x}_1,{\rm x}_2)}=4\hat{c}_{4+j}\int_{-\infty}^{\infty}d\tau_e\int_{0}^{\pi/2}d\sigma\,\frac{2\mu^2{\rm x}_{12}^{4+j}e^{j\nu\tau_e}}{({\rm x}_1^2e^{\kappa\tau_e}+{\rm x}_2^2e^{-\kappa\tau_e})^{4+j}\,\cosh^{4+j}(\mu\sigma)}\,.
}
We want to pull all the dependence on ${\rm x}_i$ out of the integral. For this purpose we change the variables $\kappa\tau_e\rightarrow\tau_e\rightarrow\tau_e+\ln\frac{{\rm x}_2}{{\rm x}_1}$, and $\mu\sigma\rightarrow\sigma$, to obtain
\eq{
G_3({\rm x}_1,{\rm x}_2,{\rm x}_3=0)=\frac{C_{S,J,j}}{{\rm x}_{12}^{2\Delta(S,J)-4-j}\,{\rm x}_1^{b_+}\,{\rm x}_2^{b_-}}\,,\qquad b_{\pm}\equiv4+j\big(1\pm{\textstyle\frac{\nu}{\kappa}}\big),
\label{3pointx12dil}
}
where
\eq{
C_{S,J,j}=\frac{8\hat{c}_{4+j}\mu}{\kappa}\int_{-\infty}^{\infty}d\tau_e\int_{0}^{\pi\mu/2}d\sigma\,\frac{e^{j\nu\tau_e/\kappa}}
{(e^{\tau_e}+e^{-\tau_e})^{4+j}\,\cosh^{4+j}\sigma}\,.
}
In the formal limit $\mu\rightarrow0$, which corresponds to the ``heavy'' operators being BPS (chiral primary), we get a vanishing structure constant \cite{Roiban:2010}, because we have an odd number of dilaton operators. Evaluating the integrals for large ${\cal S}$ and $\ell\equiv\frac{\nu}{\mu}\rightarrow0$ we find that the leading term in $C_{S,J,j}$ does not depend on ${\cal S}$
\eq{
C_{S,J,j}\,\approx\,\hat{c}_{4+j}\,\frac{4\pi\Gamma[2+\frac{j}{2}]^2}{2^{4+j}\Gamma[\frac{5+j}{2}]^2}=
\frac{\sqrt{\lambda}}{32N}\frac{\sqrt{(j+1)(j+2)(j+3)}\,\Gamma[2+\frac{j}{2}]^2}{2^j\Gamma[\frac{5+j}{2}]^2}\,,
\label{dilstrconstx12}
}
which is the same as what can be inferred from (4.15), (4.16) and (4.23) in \cite{Roiban:2010}. In the limit $j\rightarrow\infty$ we get for \eqref{dilstrconstx12}
\eq{
C_{S,J,j}\rightarrow\frac{\sqrt{\lambda}}{16N}\sqrt{j}\,e^{\,-j\ln2}\,,
}
which is expected in a case of three operators with large charges \cite{Janik:2010gc,Escobedo:2010}.

To restore the ${\rm x}_3$ dependence in \eqref{3pointx12dil} we should replace ${\rm x}_1$ with ${\rm x}_{13}$, and ${\rm x}_2$ with ${\rm x}_{23}$. The behavior of \eqref{3pointx12dil} is then the expected one for three conformal operators with dimensions
\eq{
\Delta_1=\Delta(S,J)+\frac{j\nu}{2\kappa}\,,\qquad\Delta_2=\Delta(S,J)-\frac{j\nu}{2\kappa}\,,\qquad\Delta_3=4+j\,,
}
which is consistent with charge conservation provided that $S,J\gg j$.

The four-point correlator can assume the following form
\eq{
G_4({\rm x}_1,{\rm x}_2,{\rm x}_3,{\rm x}_4)=\langle\V_{S,-J}({\rm x}_1)\V_{-S,J}({\rm x}_2)\V_{-j}({\rm x}_3)\V_j({\rm x}_4)\rangle\,.
\label{4}
}
Due to \eqref{4point}, for $S,J\gg j$, it is given by the product of two dilaton vertex operators evaluated on the solution \eqref{x12sol}. Using \eqref{2pointx12} we thus obtain\footnote{The three-point function of operators with charges $(S,-J)$, $(-S,J+j)$, and $(0,-j)$ coincides with~\eqref{3pointx12dil} provided we interchange ${\rm x}_1$ and ${\rm x}_2$.}
\eq{
G_4({\rm x}_1,{\rm x}_2,{\rm x}_3,{\rm x}_4)=\frac{(C_{S,J,j})^2}{{\rm x}_{12}^{2\Delta(S,J)}\,{\rm x}_{13}^{b_-}\,{\rm x}_{23}^{b_+}\,{\rm x}_{14}^{b_+}\,
{\rm x}_{24}^{b_-}}\,.
\label{4pointx12dil}
}
In the form of \eqref{factorization} this looks like
\eq{
G_4({\rm x}_1,{\rm x}_2,{\rm x}_3,{\rm x}_4)=\frac{\langle\V_{S,-J}({\rm x}_1)\V_{-S,J}({\rm x}_2)\V_{-j}({\rm x}_3)\rangle
\langle\V_{S,-J}({\rm x}_1)\V_{-S,J}({\rm x}_2)\V_j({\rm x}_4)\rangle}{\langle\V_{S,-J}({\rm x}_1)\V_{-S,J}({\rm x}_2)\rangle}\,.
\label{44}
}

\paragraph{Chiral primary operator-chiral primary operator}\

The three-point function with one superconformal primary scalar is again of the form~\eqref{3dilaton}, where we have substituted the dilaton vertex operator with the chiral primary one. The structure constants are then determined by the ratio $G_3({\rm x}_1,{\rm x}_2,{\rm x}_3=0)/G_2({\rm x}_1,{\rm x}_2)$ (the integral of the CPO vertex operator). Since we have assumed that ${\rm x}_1=-{\rm x}_2$, on the solution \eqref{x12sol} the chiral primary vertex operator takes its simplified form \eqref{cpovertex} (see \cite{BT:2010} for details). After integrating \eqref{cpovertex} we obtain
\al{
\frac{G_3({\rm x}_1,{\rm x}_2,{\rm x}_3=0)}{G_2({\rm x}_1,{\rm x}_2)}&=4\hat{c}_j\int_{-\infty}^{\infty}d\tau_e\int_{0}^{\pi/2}d\sigma\,\frac{2{\rm x}_{12}^je^{j\nu\tau_e}}{({\rm x}_1^2e^{\kappa\tau_e}+{\rm x}_2^2e^{-\kappa\tau_e})^j\,\cosh^j(\mu\sigma)}\nonumber\\
&\times\!\left[\frac{4\kappa^2{\rm x}_1^2{\rm x}_2^2}{({\rm x}_1^2e^{\kappa\tau_e}+{\rm x}_2^2e^{-\kappa\tau_e})^2}-\mu^2\tanh^2(\mu\sigma)\right]\!.
}
Again we want to pull all the dependence on ${\rm x}_i$ out of the integral, so we change the variables $\kappa\tau_e\rightarrow\tau_e\rightarrow\tau_e+\ln\frac{{\rm x}_2}{{\rm x}_1}$, and $\mu\sigma\rightarrow\sigma$, to get
\eq{
G_3({\rm x}_1,{\rm x}_2,{\rm x}_3=0)=\frac{C_{S,J,j}}{{\rm x}_{12}^{2\Delta(S,J)-j}\,{\rm x}_1^{j(1+\nu/\kappa)}\,{\rm x}_2^{j(1-\nu/\kappa)}}\,,
\label{3pointx12cpo}
}
where
\eq{
C_{S,J,j}=\frac{8\hat{c}_j}{\kappa\mu}\int_{-\infty}^{\infty}d\tau_e\int_{0}^{\pi\mu/2}d\sigma\,
\frac{e^{j\nu\tau_e/\kappa}}{(e^{\tau_e}+e^{-\tau_e})^j\,\cosh^j\sigma}\!\left(\frac{\kappa^2}{\cosh^2\tau_e}-\mu^2\tanh^2\sigma\right)\!.
\label{spsstrconstx12}
}
If we evaluate the integrals in the limit of small $\mu\ll\nu$, we obtain the results in subsection 4.1 of \cite{BT:2010}. In the limit of small $\nu\ll\mu$ we recover the findings in subsection 4.3 of \cite{BT:2010}.

To restore the ${\rm x}_3$ dependence in \eqref{3pointx12cpo} we should again replace ${\rm x}_1$ with ${\rm x}_{13}$, and ${\rm x}_2$ with ${\rm x}_{23}$. The behavior of \eqref{3pointx12cpo} is then the expected one for three conformal operators with dimensions
\eq{
\Delta_1=\Delta(S,J)+\frac{j\nu}{2\kappa}\,,\qquad\Delta_2=\Delta(S,J)-\frac{j\nu}{2\kappa}\,,\qquad\Delta_3=j\,.
}

Once again the four-point function can assume the form \eqref{4}. Due to \eqref{4point}, for $S,J\gg j$, it is provided by the product of two chiral primary vertex operators evaluated on the solution \eqref{x12sol}. Using \eqref{2pointx12} we thus get
\eq{
G_4({\rm x}_1,{\rm x}_2,{\rm x}_3,{\rm x}_4)=\frac{(C_{S,J,j})^2}
{{\rm x}_{12}^{2\Delta(S,J)}\,{\rm x}_{13}^{j(1-\nu/\kappa)}\,{\rm x}_{23}^{j(1+\nu/\kappa)}\,{\rm x}_{14}^{j(1+\nu/\kappa)}\,{\rm x}_{24}^{j(1-\nu/\kappa)}}\,.
\label{4pointx12cpo}
}
In the form of \eqref{factorization} this looks like \eqref{44}.

\paragraph{Dilaton-chiral primary operator}\

The corresponding three-point correlators have already been considered in the previous two paragraphs. Once more the four-point function can have the appearance of \eqref{4}. Due to \eqref{4point}, for $S,J\gg j$, it is given by the product of the dilaton operator and the chiral primary vertex operator evaluated on the solution \eqref{x12sol}. Using \eqref{2pointx12} we thus obtain two possibilities according to the order of ``light'' operators
\al{
G_4({\rm x}_1,{\rm x}_2,{\rm x}_3,{\rm x}_4)&=\frac{C_{S,J,j}^{\rm dil}\,C_{S,J,j}^{\rm CPO}}
{{\rm x}_{12}^{2\Delta(S,J)}\,{\rm x}_{13}^{b_-}\,{\rm x}_{23}^{b_+}\,{\rm x}_{14}^{j(1+\nu/\kappa)}\,{\rm x}_{24}^{j(1-\nu/\kappa)}}\,,\\
\nonumber\\
G_4({\rm x}_1,{\rm x}_2,{\rm x}_3,{\rm x}_4)&=\frac{C_{S,J,j}^{\rm CPO}\,C_{S,J,j}^{\rm dil}}
{{\rm x}_{12}^{2\Delta(S,J)}\,{\rm x}_{13}^{j(1-\nu/\kappa)}\,{\rm x}_{23}^{j(1+\nu/\kappa)}\,{\rm x}_{14}^{b_+}\,{\rm x}_{24}^{b_-}}\,.
}

\subsection{Folded string with one spin in $AdS_5$ and two equal spins in the sphere}

Let us consider examples of four-point correlators involving two non-BPS ``heavy'' operators dual to the folded string with spin $S$ in $AdS_5$ and $J_1=J_2=J/2$ in $S^5$ analyzed in Section 3. However, here we will work with generic positions ${\rm x}_1,\,{\rm x}_2$ of the form ${\rm x}_1=-{\rm x}_2$. The respective Euclidean stationary point solution is (again we choose the points ${\rm x}_1$ and ${\rm x}_2$ to lie on the $x_{0e}$-axis)
\al{\nonumber
&z=\frac{{\rm x}_{12}}{2\cosh(\kappa\tau_e)\,\cosh(\mu\sigma)}\,,\qquad x_{0e}=\frac{{\rm x}_{12}}{2}\tanh(\kappa\tau_e)+\frac12({\rm x}_1+{\rm x}_2)\,,\\ \label{x12sol'}
&x_1+ix_2=\frac{{\rm x}_{12}}{2}\frac{\tanh(\mu\sigma)}{\cosh(\kappa\tau_e)}\,e^{i\phi_1},\quad\phi_1=-i\kappa\tau_e\,,\quad
\psi=n\sigma\,,\quad\varphi_1=\varphi_2=-i\nu\tau_e\,,\\
&\kappa=\sqrt{\mu^2+\nu^2+n^2}\,,\qquad\mu=\frac{1}{\pi}\ln{\cal S}\,,\qquad{\cal S}=\frac{S}{\sqrt{\lambda}}\gg1\,,\qquad
\nu={\cal J}=\frac{J}{\sqrt{\lambda}}\,.
\nonumber
}
We denote the corresponding ``heavy'' vertex operator as $\V_{S,-J}$, with $\V_{-S,J}\equiv\V^*_{S,-J}$. In the limit of large ${\cal S}$ the two-point function of such operators is given again by \eqref{2pointx12}, where now $J$ is the sum of orbital momenta in the sphere.

The respective three-point correlators with two large spin and one BPS operators have already been calculated in Section 3, albeit for specific values of ${\rm x}_1$ and ${\rm x}_2$. Since we need 3-point functions for general ${\rm x}_1$ and ${\rm x}_2$, we will provide them below. Again, we will consider three choices for the two ``light'' operators in the four-point correlators, namely dilaton-dilaton, dilaton-CPO, and CPO-CPO.

\paragraph{Dilaton-dilaton}\

The three-point function with one dilaton assumes the following form
\eq{
G_3({\rm x}_1,{\rm x}_2,{\rm x}_3=0)=\langle\V_{S,-J}({\rm x}_1)\V_{-S,J-j}({\rm x}_2)\V_j(0)\rangle\,,\qquad S,J\gg j\,,
\label{3dilaton'}
}
where $j$ is the angular momentum of the dilaton, and we have explicitly maintained the $S^5$ angular momentum conservation. The structure constants are then determined by the ratio $G_3({\rm x}_1,{\rm x}_2,{\rm x}_3=0)/G_2({\rm x}_1,{\rm x}_2)$, which is given by the integral of the vertex operator~\eqref{dilvertex}. Evaluated on \eqref{x12sol'}, it can be expressed as
\eq{
\frac{G_3({\rm x}_1,{\rm x}_2,{\rm x}_3=0)}{G_2({\rm x}_1,{\rm x}_2)}=4\hat{c}_{4+j}\int_{-\infty}^{\infty}d\tau_e\int_{0}^{\pi/2}d\sigma\,
\frac{2(\mu^2+n^2){\rm x}_{12}^{4+j}e^{j\nu\tau_e}\cos^j(n\sigma)}{({\rm x}_1^2e^{\kappa\tau_e}+{\rm x}_2^2e^{-\kappa\tau_e})^{4+j}\,\cosh^{4+j}(\mu\sigma)}\,.
}
We want to pull all the dependence on ${\rm x}_i$ out of the integral. For this purpose we change the variables $\kappa\tau_e\rightarrow\tau_e\rightarrow\tau_e+\ln\frac{{\rm x}_2}{{\rm x}_1}$, and $\mu\sigma\rightarrow\sigma$, to obtain
\eq{
G_3({\rm x}_1,{\rm x}_2,{\rm x}_3=0)=\frac{C_{S,J,j}}{{\rm x}_{12}^{2\Delta(S,J)-4-j}\,{\rm x}_1^{b_+}\,{\rm x}_2^{b_-}}\,,\qquad b_{\pm}\equiv4+j\big(1\pm{\textstyle\frac{\nu}{\kappa}}\big),
\label{3pointx12dil'}
}
where
\eq{
C_{S,J,j}=\frac{8\hat{c}_{4+j}(\mu^2+n^2)}{\kappa\mu}\int_{-\infty}^{\infty}d\tau_e\int_{0}^{\pi\mu/2}d\sigma\,
\frac{e^{j\nu\tau_e/\kappa}\cos^j\!\big({\textstyle\frac{n\sigma}{\mu}}\big)}{(e^{\tau_e}+e^{-\tau_e})^{4+j}\,\cosh^{4+j}\sigma}\,.
}
In the formal limit $\mu\rightarrow0$ we get a non-vanishing structure constant \eqref{dilmu0}, because the string is wound around $\psi$. Evaluating the integrals for large ${\cal S}$ and $\ell_1\equiv\frac{\nu}{\mu}\rightarrow0$, while keeping $\ell_2\equiv\frac{n}{\mu}$ constant, we find that the leading term in $C_{S,J,j}$ does not depend on ${\cal S}$, but its form is complicated and devoid of physical intuition. If we further specialize to the limit of large $j$ (using saddle-point approximation), we obtain from \eqref{diljinf}-\eqref{diljinf2}
\eq{
C_{S,J,j}\rightarrow\frac{\sqrt{\lambda}}{N}\sqrt{j}\,.
}

To restore the ${\rm x}_3$ dependence in \eqref{3pointx12dil'} we should again replace ${\rm x}_1$ with ${\rm x}_{13}$, and ${\rm x}_2$ with ${\rm x}_{23}$. The behavior of \eqref{3pointx12dil'} is then the expected one for three conformal operators with dimensions
\eq{
\Delta_1=\Delta(S,J)+\frac{j\nu}{2\kappa}\,,\qquad\Delta_2=\Delta(S,J)-\frac{j\nu}{2\kappa}\,,\qquad\Delta_3=4+j\,,
}
which is consistent with charge conservation provided that $S,J\gg j$.

The four-point correlator can assume the following form
\eq{
G_4({\rm x}_1,{\rm x}_2,{\rm x}_3,{\rm x}_4)=\langle\V_{S,-J}({\rm x}_1)\V_{-S,J}({\rm x}_2)\V_{-j}({\rm x}_3)\V_j({\rm x}_4)\rangle\,.
\label{4'}
}
Due to \eqref{4point}, for $S,J\gg j$, it is given by the product of two dilaton vertex operators evaluated on the solution \eqref{x12sol'}. Using \eqref{2pointx12} we thus obtain\footnote{The three-point function of operators with charges $(S,-J)$, $(-S,J+j)$, and $(0,-j)$ coincides with~\eqref{3pointx12dil'} provided we interchange ${\rm x}_1$ and ${\rm x}_2$.}
\eq{
G_4({\rm x}_1,{\rm x}_2,{\rm x}_3,{\rm x}_4)=\frac{(C_{S,J,j})^2}{{\rm x}_{12}^{2\Delta(S,J)}\,{\rm x}_{13}^{b_-}\,{\rm x}_{23}^{b_+}\,{\rm x}_{14}^{b_+}\,
{\rm x}_{24}^{b_-}}\,.
\label{4pointx12dil'}
}
In the form of \eqref{factorization} this looks like
\eq{
G_4({\rm x}_1,{\rm x}_2,{\rm x}_3,{\rm x}_4)=\frac{\langle\V_{S,-J}({\rm x}_1)\V_{-S,J}({\rm x}_2)\V_{-j}({\rm x}_3)\rangle
\langle\V_{S,-J}({\rm x}_1)\V_{-S,J}({\rm x}_2)\V_j({\rm x}_4)\rangle}{\langle\V_{S,-J}({\rm x}_1)\V_{-S,J}({\rm x}_2)\rangle}\,.
\label{44'}
}

\paragraph{Chiral primary operator-chiral primary operator}\

The three-point function with one superconformal primary scalar is again of the form~\eqref{3dilaton'}, where we have substituted the dilaton vertex operator with the chiral primary one. The structure constants are then determined by the ratio $G_3({\rm x}_1,{\rm x}_2,{\rm x}_3=0)/G_2({\rm x}_1,{\rm x}_2)$ (the integral of the CPO vertex operator). Since we have assumed that ${\rm x}_1=-{\rm x}_2$, on the solution \eqref{x12sol'} the chiral primary vertex operator takes its simplified form \eqref{cpovertex} (see \cite{BT:2010} for details). After integrating \eqref{cpovertex} we obtain
\al{
\frac{G_3({\rm x}_1,{\rm x}_2,{\rm x}_3=0)}{G_2({\rm x}_1,{\rm x}_2)}&=4\hat{c}_j\int_{-\infty}^{\infty}d\tau_e\int_{0}^{\pi/2}d\sigma\,
\frac{2{\rm x}_{12}^j\,e^{j\nu\tau_e}\cos^j(n\sigma)}{({\rm x}_1^2e^{\kappa\tau_e}+{\rm x}_2^2e^{-\kappa\tau_e})^j\,\cosh^j(\mu\sigma)}\nonumber\\
&\times\!\left[\frac{4\kappa^2{\rm x}_1^2{\rm x}_2^2}{({\rm x}_1^2e^{\kappa\tau_e}+{\rm x}_2^2e^{-\kappa\tau_e})^2}-\mu^2\tanh^2(\mu\sigma)-n^2\right]\!.
}
Again we want to pull all the dependence on ${\rm x}_i$ out of the integral, so we change the variables $\kappa\tau_e\rightarrow\tau_e\rightarrow\tau_e+\ln\frac{{\rm x}_2}{{\rm x}_1}$, and $\mu\sigma\rightarrow\sigma$, to get
\eq{
G_3({\rm x}_1,{\rm x}_2,{\rm x}_3=0)=\frac{C_{S,J,j}}{{\rm x}_{12}^{2\Delta(S,J)-j}\,{\rm x}_1^{j(1+\nu/\kappa)}\,{\rm x}_2^{j(1-\nu/\kappa)}}\,,
\label{3pointx12cpo'}
}
where
\eq{
C_{S,J,j}=\frac{8\hat{c}_j}{\kappa\mu}\int_{-\infty}^{\infty}d\tau_e\int_{0}^{\pi\mu/2}d\sigma\,
\frac{e^{j\nu\tau_e/\kappa}\cos^j\!\big({\textstyle\frac{n\sigma}{\mu}}\big)}
{(e^{\tau_e}+e^{-\tau_e})^j\,\cosh^j\sigma}\!\left(\frac{\kappa^2}{\cosh^2\tau_e}-\mu^2\tanh^2\sigma-n^2\!\right)\!.
\label{spsstrconstx12'}
}
If we evaluate the integrals in the limit of small $\mu\ll\nu$ and $n\ll\mu$, we obtain \eqref{cpomu0l1inf}. In the limit of small $\nu\ll\mu$ and $n\ll\mu$ we recover \eqref{cpomuinfl10}. The large $j$, large ${\cal S}$ limit yields via saddle-point approximation
\eq{
C_{S,J,j}\approx\frac{\sqrt{\lambda}}{N}\frac{\ell_1}{1+\ell_2^2}\sqrt{j}\,e^{jh(\ell_1,\ell_2)},
}
where $h(\ell_1,\ell_2)$ is given by \eqref{diljinf1}.

To restore the ${\rm x}_3$ dependence in \eqref{3pointx12cpo'} we should again replace ${\rm x}_1$ with ${\rm x}_{13}$, and ${\rm x}_2$ with ${\rm x}_{23}$. The behavior of \eqref{3pointx12cpo'} is then the expected one for three conformal operators with dimensions
\eq{
\Delta_1=\Delta(S,J)+\frac{j\nu}{2\kappa}\,,\qquad\Delta_2=\Delta(S,J)-\frac{j\nu}{2\kappa}\,,\qquad\Delta_3=j\,.
}

Once again the four-point function can assume the form \eqref{4'}. Due to \eqref{4point}, for $S,J\gg j$, it is provided by the product of two chiral primary vertex operators evaluated on the solution \eqref{x12sol'}. Using \eqref{2pointx12} we thus get
\eq{
G_4({\rm x}_1,{\rm x}_2,{\rm x}_3,{\rm x}_4)=\frac{(C_{S,J,j})^2}
{{\rm x}_{12}^{2\Delta(S,J)}\,{\rm x}_{13}^{j(1-\nu/\kappa)}\,{\rm x}_{23}^{j(1+\nu/\kappa)}\,{\rm x}_{14}^{j(1+\nu/\kappa)}\,{\rm x}_{24}^{j(1-\nu/\kappa)}}\,.
\label{4pointx12cpo'}
}
In the form of \eqref{factorization} this looks like \eqref{44'}.

\paragraph{Dilaton-chiral primary operator}\

The corresponding three-point correlators have already been considered in the previous two paragraphs. Once more the four-point function can have the appearance of \eqref{4'}. Due to \eqref{4point}, for $S,J\gg j$, it is given by the product of the dilaton operator and the chiral primary vertex operator evaluated on the solution \eqref{x12sol'}. Using \eqref{2pointx12} we thus obtain two possibilities according to the order of ``light'' operators
\al{
G_4({\rm x}_1,{\rm x}_2,{\rm x}_3,{\rm x}_4)&=\frac{C_{S,J,j}^{\rm dil}\,C_{S,J,j}^{\rm CPO}}
{{\rm x}_{12}^{2\Delta(S,J)}\,{\rm x}_{13}^{b_-}\,{\rm x}_{23}^{b_+}\,{\rm x}_{14}^{j(1+\nu/\kappa)}\,{\rm x}_{24}^{j(1-\nu/\kappa)}}\,,\\
\nonumber\\
G_4({\rm x}_1,{\rm x}_2,{\rm x}_3,{\rm x}_4)&=\frac{C_{S,J,j}^{\rm CPO}\,C_{S,J,j}^{\rm dil}}
{{\rm x}_{12}^{2\Delta(S,J)}\,{\rm x}_{13}^{j(1-\nu/\kappa)}\,{\rm x}_{23}^{j(1+\nu/\kappa)}\,{\rm x}_{14}^{b_+}\,{\rm x}_{24}^{b_-}}\,.
}

\sect{Conclusion}

The AdS/CFT correspondence passed through many controversial developments over the last decade. The holographic
conjecture has been tested in many cases and impressive results about anomalous dimensions of the gauge theory
operators, integrable structures, etc., and crucial properties of various gauge theories at strong coupling have been
established. The main challenge ahead is to find efficient methods for calculation of the correlation functions.

While discovering a semiclassical trajectory controlling the leading contribution to the three-point correlator of ``heavy'' operators is so far an unsolved problem \cite{Janik:2010gc}, we have seen that one can use the trajectory for the correlation function of two ``heavy'' operators, which is straightforward to find \cite{Tseytlin:2003,Buchbinder:2010vw}, to compute the behavior of a correlator containing two ``heavy'' and one ``light'' states at strong coupling \cite{Zarembo:2010,Costa:2010}\footnote{Such considerations were initiated in \cite{Janik:2010gc}.}. The approach based on insertion of vertex operators was put forward in \cite{Roiban:2010}, where the authors also suggested that the same method applies to higher $n$-point correlation functions with 2 ``heavy'' and $n-2$ ``light'' operators. Namely, the semiclassical expression for $n$-point correlator should be given by a product of ``light'' vertex operators calculated on the worldsheet surface determined by the ``heavy'' operator insertions.

In the present paper we considered string theory on $\axs$ and computed three-point correlation functions of two ``heavy'' (string) and one ``light'' (supergravity) states at strong coupling, applying the ideas of \cite{Roiban:2010} for calculation of correlation functions using vertex operators for the corresponding states. We examined the method in the case of a folded string solution with three spins (one in AdS and two equal ones in the sphere), which generalizes the solution used in \cite{Roiban:2010}. We provided a number of limiting cases, which illuminate the physical motivation behind the calculations. We also computed several four-point correlators for folded string solutions saturating dilaton and chiral primary ``light'' operators, and extending the results presented in \cite{BT:2010}.

Very recently there has been development in the semiclassical calculation of two- and three-point correlation functions of operators dual to open string and brane states (giant gravitons in particular) \cite{Bak:2011,Bissi:2011}. One of the important directions for future research in this field is the computation of correlators of wrapped branes on vanishing cycles.

\section*{Acknowledgments}
The authors would like to thank H. Dimov for valuable discussions and careful reading of the paper.
This work was supported in part by the Austrian Research Funds FWF P22000 and I192, NSFB VU-F-201/06 and DO 02-257.


\end{document}